\documentstyle[12pt,epsf]{article}
\newlength{\mytopmargin}
\newlength{\myleftmargin}
\def\zz{\relax\hbox{\small \sf Z\kern-.4em Z}}

\newcommand{\ml}{\langle}
\newcommand{\mg}{\rangle}

\renewcommand{\theequation}{\thesection.\arabic{equation}}

\begin{document}
\vspace{4cm}
\noindent
\begin{center}{\bf CORRELATIONS FOR THE CIRCULAR DYSON BROWNIAN\\
 MOTION MODEL WITH POISSON INITIAL CONDITIONS}
\end{center}
\vspace{5mm}

\noindent
\begin{center}
 P.J.~Forrester\footnote{ Department of Mathematics, University of Melbourne, Parkville, Victoria 
3052, Australia; email: matpjf@maths.mu.oz.au; supported by the ARC}
and
T.~Nagao\footnote {Department of Physics, Faculty of Science, Osaka University,
Toyonaka, Osaka 560, Japan;
email: nagao@sphinx.phys.sci.osaka-u.ac.jp}
\end{center}
\vspace{.5cm}

\small
\begin{quote}
The circular Dyson Brownian motion model refers to the stochastic
dynamics of the  log-gas on a circle. It also specifies
the eigenvalues of certain parameter-dependent ensembles of unitary
random matrices. This model is considered with the initial condition
that the particles are non-interacting (Poisson statistics). 
Jack polynomial theory is used to derive a simple
exact expression for the density-density correlation with the position
of one
particle specified in the initial state, and the position of one
particle specified at time $\tau$, valid for all $\beta > 0$.
 The same correlation with two
particles specified in the initial state is also derived exactly, and
some special cases of the theoretical correlations are illustrated
by comparison with the empirical correlations calculated from the
eigenvalues of certain parameter-dependent Gaussian random matrices.
Application to fluctuation formulas for time displaced linear
statistics in made.
\end{quote}

\vspace{.5cm}
\noindent
\section{Introduction}
\setcounter{equation}{0}
The Dyson Brownian motion model \cite{Dy63} refers to the overdamped Brownian
dynamics of the one-dimensional log-gas. This dynamics is specified by
the Fokker-Planck equation
\begin{equation}\label{fp}
{\partial p \over \partial \tau} ={\cal L}p \qquad \mbox{where} \quad
{\cal L} = \sum_{j=1}^N {\partial  \over \partial x_j}
\left (   {\partial W \over \partial x_j}
+\beta^{-1} {\partial  \over \partial x_j} \right )
\end{equation}
with the particular potential
\begin{equation}\label{w1}
W = {1 \over 2}
\sum_{j=1}^N x_j^2 - \sum_{1 \le j < k \le N} \log | x_k - x_j|,
\end{equation}
or its periodic version
\begin{equation}\label{w2}
W =  - \sum_{1 \le j < k \le N}
\log |e^{2 \pi i x_k/L} - e^{2 \pi i x_j/L}|.
\end{equation}

The formulation of this model has its origin in the theory of parameter
dependent Gaussian random matrices. We recall that a random matrix $H$
is of this type if its joint distribution of elements is proportional to
\begin{equation}\label{pdfg}
\exp \Big ( - \beta {\rm Tr}(H - e^{-\tau} H^{(0)})^2/(2(1 - e^{-2\tau}))
\Big ),
\end{equation}
where $\beta =1,2$ or 4 according to $H$ being symmetric, Hermitian or
self dual real quaternion respectively. This means that each part (real or
imaginary) of each independent element of $H$ is chosen with a
particular Gaussian distribution. The matrix $H^{(0)}$ is some prescribed
(random) matrix. Note that as $\tau \to 0$, $H = H^{(0)}$, while as
$\tau \to \infty$ (\ref{pdfg}) reduces to $\exp(-\beta{\rm Tr} H^2/2)$.
With $\beta =1$, 2 and 4 this latter distribution
specifies the Gaussian Orthogonal, Unitary and Symplectic
Ensembles (GOE, GUE, GSE)  respectively. 

Let $p=p(x_1,\dots,x_N;\tau)$ denote the eigenvalue 
probability density function (p.d.f.) corresponding
to (\ref{pdfg}). By using second order perturbation theory, Dyson
\cite{Dy63} showed that $p$ satisfies the Fokker-Planck equation
(\ref{fp}) with $W$ given by (\ref{w1}). This equation must be solved
subject to the initial condition that $p$ corresponds to the
eigenvalue p.d.f.~of $H$ at $\tau = 0$. Dyson also formulated a
parameter dependent theory of random unitary matrices from the
so-called circular ensembles. Here the situation is more abstract in
the sense that no explicit construction of such unitary random matrices
is known. Nonetheless, in the framework of the abstract formulation it was
shown that the corresponding eigenvalue p.d.f.~(eigenvalues $e^{2 \pi i
x_j/L}$, $0 \le x_j < L$) satisfies the Fokker-Planck equation
(\ref{fp}) with $W$ given by (\ref{w2}). It turns out that upon
introducing the scaling \cite{Fo96z}
\begin{equation}\label{sc}
x \mapsto \pi \rho x / \sqrt{2N}, \qquad \tau \mapsto (\pi \rho)^2 \tau/
(2N)
\end{equation}
into the Fokker-Planck equation with $W$ given by (\ref{w1}), and taking
the $N \to \infty$ limit, the results obtained for the correlation
functions and other observable quantities are identical to those
obtained with the choice of $W$ (\ref{w2}). In fact for correlations
over one or two eigenvalue spacings in the bulk of the spectrum,
the agreement already becomes apparant for matrix dimensions $N = 11$
(see e.g.~\cite{FN97}).

Our specific interest in this paper is the exact calculation of certain
correlation functions in the Fokker-Planck equation (\ref{fp})
with $W$ given by (\ref{w2}) and subject to the initial condition
$p(x_1,\dots,x_N;0) = {1 \over L^N}$. This initial condition corresponds
to non-interacting particles; in the applied random matrix literature 
it is referred to as Poisson initial conditions, which is consistent since a
gas of non-interacting particles exhibits Poisson statistics.
Thus for the random matrix couplings $\beta =1,2$ and 4
and $N$ large, the Fokker-Planck
equation with this initial condition describes the transition
between an eigenvalue p.d.f.~with Poisson statistics to the
eigenvalue p.d.f.~of the appropriate Gaussian ensemble. Such a transition
is relevant to the description of the statistical properties of the
eigenvalue spectrum in a quantum system which is initially integrable,
but becomes chaotic as a parameter (usually identified with $\tau^{1/2}$;
see e.g.~\cite{BR94}) is varied. Because of this application, this problem
has received a lot of recent attention \cite{Ha92,BH96,Pa95,Gu96,KS98}.

In the next section we will specify the particular correlation functions to 
be calculated (density-density correlation with the position
of $n$ particles specified in the initial state, and the position of one
particle specified at `time' $\tau$), and then proceed to calculate
this correlation in the case $n=1$ for general $\beta$. 
Comparison with the empirical evaluation of this correlation for
$\beta = 1$ obtained from 
numerically generated parameter dependent random matrices is also
made. In Section 3 we consider the same correlation with $n=2$, and
provide its exact value for $\beta = 2$ and 4, as well as all
$0 < \beta < 2$. A
discussion of these results, including asymptotic properties and
their relationship to fluctuation formulas for linear
statistics, is given in Section 4.

\section{Density-density correlation}
\setcounter{equation}{0}
\subsection{Formalism}
In general the
correlation functions for the
Brownian motion described by (\ref{fp}) 
can be specified in terms of the Green
function $G(x_1^{(0)},\dots,x_N^{(0)};x_1,\dots,x_N;\tau)$, which is
by definition the solution of (\ref{fp}) subject to the initial
condition
\begin{equation}\label{df}
p(x_1,\dots,x_n;0) = \prod_{j=1}^N \delta(x_j - x_j^{(0)}).
\end{equation}
For a general initial condition with p.d.f.~$f(x_1^{(0)},\dots,x_N^{(0)})$
(assumed symmetric) the particular density-density correlation
$\rho_{(n,1)}^T(x_1^{(0)},\dots,x_n^{(0)};x;\tau)$ is given in terms
of $G$ by
\begin{eqnarray}\label{dd}\lefteqn{
\rho_{(n,1)}^T(x_1^{(0)},\dots,x_n^{(0)};x;\tau)  = 
N(N-1) \cdots (N-n+1) \int_I dx_{n+1}^{(0)} \cdots \int_I dx_N^{(0)} \,
f(x_1^{(0)},\dots,x_N^{(0)})} \nonumber \\
&& \times \int_I dx_1^{(1)} \cdots \int_I dx_N^{(1)} \,
\Big ( \sum_{j=1}^N \delta (x - x_j^{(1)}) \Big )
G(x_1^{(0)},\dots,x_N^{(0)};x_1,\dots,x_N;\tau) \qquad \qquad \nonumber \\&&
- \rho_{(n)}(x_1^{(0)},\dots,x_n^{(0)};0) \rho_{(1)}(x;\tau).
\end{eqnarray}
Note that $\rho_{(n,1)}^T(x_1^{(0)},\dots,x_n^{(0)};x;\tau)/
\rho_{(n)}(x_1^{(0)},\dots,x_n^{(0)};0) + \rho_{(1)}(x;\tau)$
represents the density at position $x$ after time $\tau$, given
that there are particles at $x_1^{(0)}, \dots, x_n^{(0)}$ initially.

To obtain a useful expression for the Green function, one uses the
general fact \cite{Ri84} that after conjugation with $e^{-\beta W}$
the Fokker-Planck operator ${\cal L}$ transforms into a Hermitian
operator. In fact (see e.g.~\cite{Fo98}) for the choices of $W$
(\ref{w1}) and (\ref{w2}) one has
\begin{equation}\label{lh}
e^{\beta W/2} {\cal L} e^{-\beta W/2} = - {1 \over \beta} (H - E_0),
\end{equation}
where $H$ is the Schr\"odinger operator for a quantum mechanical system
with one and two body interactions only, and $E_0$ is the corresponding
ground state energy. Explicitly, for $W$ given by (\ref{w1})
\begin{equation}\label{csm}
H = - \sum_{j=1}^N {\partial^2 \over \partial x_j^2}
+ \beta (\beta/2 - 1) \Big ( {\pi \over L} \Big )^2
\sum_{1 \le j < k \le N}
{1 \over \sin^2 \pi (x_k - x_j)/L},
\end{equation}
which is an example of the Calogero-Sutherland quantum many body
system (quantum particles with $1/r^2$ pair interaction).
Now in general for the imaginary time Schr\"odinger
equation
\begin{equation}\label{it}
{\partial \psi \over \partial \tau} = 
- {1 \over \beta} (H - E_0) \psi 
\end{equation}
in which $H$ has
 a complete set
of orthogonal eigenfunctions $\{\psi_\kappa(x)\}_\kappa$
(here $x:= (x_1,\dots,x_N)$ and $\kappa$ represents an
$n$-tuple of ordered integers $\kappa_1 \ge \kappa_2 \ge \dots
\ge \kappa_N$) with corresponding eigenvalues $\{E_\kappa\}_\kappa$,
the method of separation of variables gives that the Green
function solution is
\begin{equation}\label{gfs}
G^{(H)}(x_1^{(0)},\dots,x_N^{(0)};x_1,\dots,x_N;\tau)
= \sum_{\kappa} {\psi_\kappa^*(x^{(0)}) \psi_\kappa(x) \over
\ml \psi_\kappa | \psi_\kappa \mg} e^{-(E_\kappa - E_0)\tau/\beta}.
\end{equation}
In (\ref{gfs})
\begin{equation}\label{gfs1}
\ml \psi_\kappa | \psi_\kappa \mg := \int_Idx_1 \cdots
\int_I dx_N |\psi_\kappa(x)|^2,
\end{equation}
and ${}^*$ denotes complex conjugation. 

Since, according to (\ref{lh}), the Green function solution 
$G(x^{(0)};x;\tau)$ of the Fokker-Planck equation
(\ref{fp}) is related to the Green function solution 
$G^{(H)}(x^{(0)};x;\tau)$ of
(\ref{it}) by
\begin{equation}\label{gfr}
G(x^{(0)};x;\tau) = {e^{-\beta W(x_1,\dots,x_N)/2} \over
e^{-\beta W(x_1^{(0)},\dots,x_N^{(0})/2}} G^{(H)}(x^{(0)};x;\tau),
\end{equation}
$G(x^{(0)};x;\tau)$ is determined to the extent that the quantities
in (\ref{gfs}) are known. Independent of such explicit knowledge,
substituting (\ref{gfs}) in (\ref{gfr}) and substituting the result
in (\ref{dd}) gives
\begin{eqnarray}\label{dd1}\lefteqn{
\rho_{(n,1)}^T(x_1^{(0)},\dots,x_n^{(0)};x;\tau)  = 
N(N-1) \cdots (N-n+1)} \\&&
\times \sum_{\kappa, \kappa \ne 0}
{A_\kappa(x_1^{(0)},\dots,x_n^{(0)})
\ml e^{-\beta W /2} | \sum_{j=1}^N \delta (x - x_j) | \psi_\kappa \mg \over
\ml \psi_\kappa | \psi_\kappa \mg} e^{-(E_\kappa - E_0)\tau/\beta}
\end{eqnarray}
where
\begin{eqnarray}
A_\kappa(x_1^{(0)},\dots,x_n^{(0)}) & = &
\int_I dx_{n+1}^{(0)} \cdots \int_I dx_N^{(0)} \,
f(x_1^{(0)},\dots,x_n^{(0)}) \psi_\kappa(x^{(0)})  e^{\beta W /2}
\label{a1} \\
\ml e^{-\beta W /2} | \sum_{j=1}^N \delta (x - x_j) | \psi_\kappa \mg
& = & N \int_I dx_2 \cdots \int_I dx_N \, e^{-\beta W(x) /2}
\psi_\kappa(x). \label{d1}
\end{eqnarray}

\subsection{Explicit formulas}
For the Schr\"odinger operator (\ref{csm}), it is known
\cite{Su72,Fo94} that the (unnormalized) eigenfunctions can be
written in the form
\begin{equation}\label{fact}
\psi_\kappa(x) = e^{-\beta W/2} P_\kappa^{(2/\beta)}(z), \qquad
z := e^{2 \pi i x/L},
\end{equation}
where $\kappa = (\kappa_1,\dots,\kappa_N)$
forms a partition and thus has all parts non-negative and ordered
as specified below (\ref{it}), and $P_\kappa^{(2/\beta)}(z)$ is a
particular polynomial known as the Jack polynomial. Each Jack
polynomial has the special structure
\begin{equation}\label{fact1}
P_\kappa^{(2/\beta)}(z) = m_\kappa + \sum_{\mu < \kappa}
b_{\kappa \mu} m_\mu
\end{equation}
where $m_\kappa$ refers to the monomial symmetric function
(in the variables $z_1,\dots,z_N$)
corresponding to the partition $\kappa$, the $b_{\kappa \mu}$ are
coefficients, while $\mu < \kappa$ refers to the dominance ordering
of partitions.  For $\kappa$ possessing negative parts, (\ref{fact1})
has no immediate meaning. In such cases the Jack polynomials are defined
by the relation
\begin{equation}\label{fact2}
P_{\kappa - c}^{(2/\beta)}(z) = \Big ( \prod_{l=1}^N z_l^{-c} \Big )
P_\kappa^{(2/\beta)}(z),
\end{equation}
where $\kappa$ is a partition while $c:=(c,\dots,c)$, $c \in \zz^+$.
Also, for general $\kappa$, the energy eigenvalues are given by
\begin{equation}\label{fact3}
E_\kappa - E_0 = \Big ({2 \pi \over L} \Big )^2
\Big ( \sum_{j=1}^N \kappa_j^2 + {\beta \over 2}
\sum_{j=1}^N \kappa_j(N - 2j + 1) \Big ).
\end{equation}

Although (\ref{fact1}) is a very special property associated with
(\ref{csm}), the true utility of Jack polynomial theory in regard
to computing correlation functions lies with our knowledge
of the integrals (\ref{gfs1}) and (\ref{d1}) (see e.g.~\cite{Fo98}
and references therein).
First consider the normalization integral (\ref{gfs1}). In the
case of the ground state $(\kappa = 0)$ we have
\begin{eqnarray}
\ml \psi_0 | \psi_0 \mg & := &
\int_0^L dx_1 \cdots \int_0^L dx_N \,
\prod_{1 \le j < k \le N} |e^{2 \pi i x_j/L} - e^{2 \pi i x_k /L}
|^\beta \nonumber
\\ & = & L^N {\Gamma(\beta N /2 + 1) \over (\Gamma(\beta/2 + 1))^N}.
\end{eqnarray}
This can be conveniently factored out of the normalization integral for
general $\kappa$. In addition, the general $\kappa$ case involves
the generalized factorial 
\begin{equation}
[u]_\kappa^{(2/\beta)} := \prod_{j=1}^N
{\Gamma(u - {\beta \over 2}(j-1) + \kappa_j) \over
\Gamma(u - {\beta \over 2}(j-1))},
\end{equation}
the magnitude of the partition,
\begin{equation}
| \kappa | := \sum_{j=1}^N \kappa_j,
\end{equation}
the Jack polynomial which each variable set equal to unity
$P_\kappa^{(2/\beta)}(1^N)$, 
and the quantity
\begin{equation}\label{kpr}
d_\kappa':= \prod_{(i,j) \in \kappa}
\Big ( \kappa_j' - i + {2 \over \beta}(\kappa_i - j + 1) \Big ).
\end{equation}
In (\ref{kpr}) it is necessary that $\kappa$ be a partition
The product is over all squares in the diagram of $\kappa$,
and $\kappa'$ refers to the partition conjugate to $\kappa$.
In terms of this notation, for $\kappa$ a partition we have 
\begin{equation}\label{norm}
\ml \psi_\kappa | \psi_\kappa \mg =
\ml \psi_0 | \psi_0 \mg {(\beta/2)^{|\kappa|} d_\kappa'
P_\kappa^{(2/\beta)}(1^N) \over
[ {\beta \over 2}(N - 1) + 1]_\kappa^{(2/\beta)}}.
\end{equation}
The normalization of the states in which $\kappa$ possesses
negative parts and so does not form a partition are
reducible to the latter case by the simple formula
\begin{equation}
\Big \ml \Big ( \prod_{l=1}^N z_l^{-c} \Big ) \psi_\kappa \Big |
\Big ( \prod_{l=1}^N z_l^{-c} \Big ) \psi_\kappa \Big \mg =
\ml \psi_\kappa | \psi_\kappa \mg.
\end{equation}

We turn now to the integral (\ref{d1}). This quantity has the
evaluation
\begin{equation}\label{deval}
\ml e^{-\beta W /2} | \sum_{j=1}^N \delta (x - x_j) | \psi_\kappa \mg
= e^{2 \pi i x |\kappa|/L} \ml \psi_0 | \psi_0 \mg
{|\kappa| (\kappa_1 - 1)! \over L} P_\kappa^{(2/\beta)}(1^N) 
{\prod_{j=2}^{\ell(\kappa)} \Big ( - {\beta \over 2}(j-1) \Big )_{\kappa_j}
\over [{\beta \over 2}(N - 1) + 1]_\kappa^{(2/\beta)}},
\end{equation}
where $\ell(\kappa)$ denotes the length of the partition $\kappa$
(i.e.~the number of non-zero parts) and
\begin{equation}
(a)_n := a(a+1) \cdots (a+n-1).
\end{equation}
For the $n$-tuples of non-positive integers
\begin{equation}
- \bar{\kappa} := (-\kappa_N, - \kappa_{N-1}, \dots, - \kappa_1)
\end{equation}
we have the simple formula
\begin{equation}\label{sim}
\Big \ml \psi_0 \Big | \sum_{j=1}^N \delta(x-x_j) \Big | \psi_{- \bar{\kappa}}
\Big \mg =
\Big \ml \psi_0 \Big | \sum_{j=1}^N \delta(x-x_j) \Big | \psi_\kappa \Big \mg^*,
\end{equation}
and thus (\ref{deval}) gives the evaluation. Furthermore, for $\beta$
rational, we have the result \cite{Fo95} that (\ref{d1}) is non-zero only if all
parts of $\kappa$ are non-negative or all parts are non-positive.
Hence, if we restrict attention to $\beta$ rational (which we will do
henceforth), the results (\ref{deval}) and (\ref{sim}) suffice.
In this regard we note also that it is a simple result that
\begin{equation}
\ml \psi_{-\bar{\kappa}}|\psi_{-\bar{\kappa}} \mg =
\ml \psi_\kappa | \psi_\kappa \mg,
\end{equation}
which when combined with (\ref{sim}) implies that we can restrict 
the summation in (\ref{dd1}) to partitions provided twice the
real part is taken.

Finally we come to consider the integral $A_\kappa$
defined by (\ref{a1}). For the case
under investigation, this reads
$$
A_\kappa(x_1^{(0)},\dots,x_n^{(0)}) = {1 \over L^N}
\int_0^L dx_{n+1}^{(0)} \cdots \int_0^L dx_N^{(0)}
\Big ( P_\kappa^{(2/\beta)}(z_1^{(0)},\dots,z_N^{(0)}) \Big )^*,
$$
which is proportional to the term independent of
$z_{n+1}^{(0)},\dots,z_N^{(0)}$ in the power series expansion
of $P_\kappa^{(2/\beta)}(z^{(0)})$.  
Since we have now reduced the cases to be
considered down to those for which $\kappa$ is a partition, 
this term can be determined by setting $z_{n+1}^{(0)} =
\cdots = z_N^{(0)} = 0$ and we
therefore have
\begin{eqnarray}\label{a2}
A_\kappa(x_1^{(0)},\dots,x_n^{(0)})  & = &
L^{N-n} \Big (  P_\kappa^{(2/\beta)}(z_1^{(0)},\dots,z_n^{(0)},0,
\dots,0) \Big )^*  \nonumber \\
& = & \left \{ 
\begin{array}{ll}
L^{N-n} \Big (  P_\kappa^{(2/\beta)}(z_1^{(0)},\dots,z_n^{(0)})
\Big )^* & \ell({\kappa}) \le n, \\
0 & {\rm otherwise}, \end{array} \right.
\end{eqnarray}
where the first expression in the second equality follows from the
fact that the coefficients $b_{\kappa \mu}$ in (\ref{fact1}) are
independent of $N$.

Substituting (\ref{norm}), (\ref{deval}) and (\ref{a2})
into (\ref{dd1}) and taking twice the real part we obtain
the formula
\begin{eqnarray}\label{fin}
\lefteqn{\rho_{(n,1)}^T(x_1^{(0)},\dots,x_n^{(0)};x;\tau) = } \nonumber 
\\ &&
{N(N-1) \cdots (N-n+1) \over L^n} 2 {\rm Re}
\sum_{{\rm partitions} \, \kappa \atop \kappa \ne 0, \,
\ell(\kappa) \le n} u_\kappa
\Big ( P_\kappa^{(2/\beta)}(z_1^{(0)},\dots,z_n^{(0)}) \Big )^*
e^{2 \pi i |\kappa| x / L} e^{-\tau (E_\kappa - E_0)/\beta} \nonumber 
\\
\end{eqnarray}
where
\begin{equation}\label{u}
u_\kappa := {1 \over L}
 {|\kappa| (\kappa_1 - 1)! \over d_\kappa' (\beta / 2)^{|\kappa|}}
\prod_{j=2}^{\ell(\kappa)} \Big (-{\beta \over 2} (j-1) \Big )_{\kappa_j},
\end{equation}
valid for $\beta$ rational at least. In fact this result must be valid
for all $\beta > 0$, by continuity of $\rho_{(n,1)}^T$ in $\beta$. 

\subsection{The case $n=1$}
In the case $n=1$ the only partitions contributing to (\ref{fin})
are $(\kappa_1,0,\dots,0)$ and we have
\begin{equation}
P_\kappa^{(2/\beta)}(z_1^{(0)}) = \Big ( z_1^{(0)} \Big )^{\kappa_1},
\qquad E_\kappa - E_0 = \Big ({2 \pi \over L} \Big )^2
\Big ( \kappa_1^2 + {\beta \over 2}(N - 1) \kappa_1 \Big ), \qquad
u_\kappa = 1.
\end{equation}
Thus (\ref{fin}) reduces to the remarkably simple formula
\begin{equation}\label{rms}
\rho_{(1,1)}^T(x_1^{(0)};x;\tau) =
{2 (N-1) \rho  \over N L} \sum_{\kappa_1 = 0}^\infty
e^{-(2 \pi / L)^2 (\kappa_1^2 + (\beta / 2)(N - 1) \kappa_1) \tau /
\beta} \cos 2 \pi \kappa_1(x - x_1^{(0)})/L
\end{equation}
where $\rho := N/L$.
In the thermodynamic limit $N,L \to \infty$,
$\rho$ fixed, (\ref{rms}) being a Riemann sum tends to the definite
integral
\begin{equation}\label{rms1}
\rho_{(1,1)}^T(x_1^{(0)};x;\tau) =
2 \rho^2 \int_0^\infty e^{-(2 \pi \rho)^2 (s^2 + (\beta/2)s) \tau/\beta}
\cos 2 \pi s \rho (x - x_1^{(0)}) \, ds.
\end{equation}

As mentioned in the Introduction, the correlation function given
analytically by (\ref{rms}) with $\beta =1,2$ or 4 can be calculated
empirically using parameter-dependent random matrices. For definiteness
consider the case $\beta = 1$. We construct initially a 
diagonal random matrix $H^{(0)}={\rm diag}(h_{11}^{(0)},
\dots,h_{NN}^{(0)})$ with each diagonal entry chosen from a
Gaussian distribution with mean zero and standard deviation 
$N (2 \pi)^{-1/2}$. This standard deviation is chosen so that the
eigenvalue density, which equals $N$ times the Gaussian distribution for
a diagonal element, is unity at the origin. Then, following the
prescription (\ref{pdfg}) with $\beta = 1$ and the scaling (\ref{sc}),
we construct a real symmetric random matrix in which the diagonal entries
have mean $e^{-t} h_{jj}^{(0)}$ and standard deviation
$(\sqrt{2N}/\pi)(1 - e^{-2t})^{1/2}$, $t:= \pi^2 \tau /(2N)$,
while the independent off diagonal elements (upper triangular elements
say) have mean zero and standard deviation equal to $1/\sqrt{2}$ of that of
the diagonal elements. Note in particular that the factor of
$\sqrt{2N}/ \pi$ in the standard deviation is chosen so that at
the centre of the spectrum the theoretical density remains equal to
unity. 

\begin{figure}
\epsfxsize=10cm
\centerline{\epsfbox{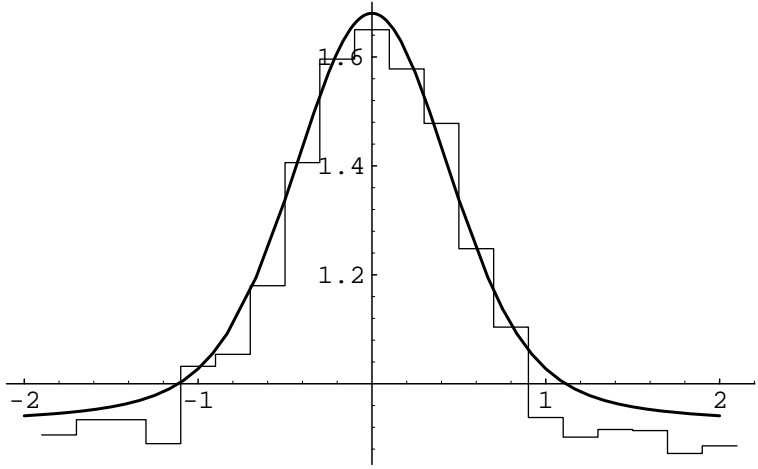}}
\caption{\label{f.1} Comparison between the empirical density for
2,500
Gaussian real symmetric parameter-dependent random matrices of
dimension $13 \times 13$ with
$t = .025$ and a zero eigenvalue initially, and
the theoretical value computed from (\ref{rms})
with $\beta=1$. }
\end{figure}

We choose a specific value of $N$ and $t$,
and numerically generate many such random matrices with a fixed initial
diagonal entry $h_{11}^{(0)} = 0$, together with
their eigenvalues. The corresponding empirical eigenvalue
 density in the neighbourhood of the origin can be
computed and compared directly against the theoretical result (\ref{rms})
with $\rho = 1$, $\beta = 1$
and the equilibrium density
$\rho = 1$ added. The results of such a calculation are given in
Figure \ref{f.1}. We see that the empirical and theoretical results
agree to statistical accuracy.

\vspace{.5cm}

\section{The correlation $\rho_{(2,1)}^T(x_1^{(0)},x_2^{(0)};x;\tau)$}
\setcounter{equation}{0}
In this section we will consider the correlation (\ref{fin})
with $n=2$. Explicit formulas are possible in this case because of
the formula \cite{Ya92}
\begin{equation}\label{3.1}
P_{(\kappa_1,\kappa_2)}^{(2/\beta)}(z_1^{(0)},z_2^{(0)}) =
{2^{\kappa_1 - \kappa_2} \over a_{\kappa_1 - \kappa_2}}
(z_1^{(0)} z_2^{(0)})^{(\kappa_1 + \kappa_2)/2}
P_{\kappa_1 - \kappa_2}^{(\gamma, \gamma)}\Big (
{1 \over 2} (z_1^{(0)} + z_2^{(0)}) (z_1^{(0)} z_2^{(0)})^{-1/2} \Big ),
\end{equation}
where on the r.h.s.~$P_n^{(\alpha,\beta)}(x)$ denotes the Jacobi polynomial,
$\gamma := (\beta - 1)/2$ and
\begin{equation}\label{an}
a_n := \Big ( {n + \gamma \atop n} \Big )
{(n + 2 \gamma + 1)_n \over (\gamma + 1)_n } 2^{-n}.
\end{equation}
Of particular relevance is the asymptotic form of (\ref{3.1})
in the limit $\kappa_1, \kappa_2,L \to \infty$, $\rho$ fixed.
The leading behaviour of the term (\ref{an}) is determined using
Stirling's formula, while the Jack polynomial is estimated by
noting that
\begin{equation}
{1 \over 2}(z_1^{(0)} + z_2^{(0)}) (z_1^{(0)} z_2^{(0)})^{-1/2} =
\cos \pi (x_1^{(0)} - x_2^{(0)})/L,
\end{equation}
and making use of the asymptotic formula
\begin{equation}
P_n^{(\gamma,\gamma)}(\cos \theta)
\sim \Big ( {2 \over \theta} \Big )^\gamma J_\gamma(n \theta),
\end{equation}
where $J_\gamma$ denotes the Bessel function. 
Thus we have
\begin{eqnarray}\label{pas}\lefteqn{
P_{(\kappa_1,\kappa_2)}^{(2/\beta)}(z_1^{(0)},z_2^{(0)})  \sim
N^{\beta / 2} \Big ( \pi (s_1 - s_2) \Big )^{1/2}} \nonumber \\&& \times
e^{\pi i (s_1 + s_2) \rho (x_1^{(0)} + x_2^{(0)})}
{1 \over (2 \pi \rho (x_1^{(0)} - x_2^{(0)}))^{(\beta - 1)/2}}
J_{(\beta - 1)/2}\Big ((\kappa_1 - \kappa_2) \pi 
\rho (x_1^{(0)} - x_2^{(0)})\Big )
\end{eqnarray}
where $s_1 := \kappa_1/N$, $s_2 := \kappa_2/N$.

Due to the factor $(-\beta/2)_{\kappa_2}$ in (\ref{u}),
the cases for which $\beta$ is even require special consideration
since then $\kappa_2$ is restricted to the range 0 to $\beta/2-1$ for
a non-zero contribution. In fact, due to cancellation effects within
this range, each even value of $\beta$ must be considered separately.
We will consider in detail the cases $\beta = 2$ and 4.

\subsection{$\beta = 2$}
In this case we see from (\ref{u}) and (\ref{kpr}) that with
$\kappa_2 =0$, $u_\kappa = 1/L$ while for $\kappa_2 = 1$, $u_\kappa = -1/L$.
This fact, together with (\ref{pas}) and (\ref{fact3}) shows
\begin{eqnarray}\lefteqn{
\sum_{\kappa_2 = 0}^1 u_\kappa
\Big ( P_{(\kappa_1,\kappa_2)}(z_1^{(0)},z_2^{(0)}) \Big )^*
e^{2 \pi i |\kappa| x/L} e^{-\tau (E_\kappa - E_0)/\beta}
\sim {\rho \pi^{1/2} \over N}
e^{-\pi i s_1 \rho (\bar{x}_1^{(0)}+(\bar{x}_2^{(0)})}
} \nonumber \\&&
\times
e^{-(2 \pi \rho)^2 (s_1^2 + s_1)\tau / 2} 
\bigg \{ {\partial \over \partial s_1} \Big [
\bigg ( { s_1 \over 2 \pi \rho (x_1^{(0)} - x_2^{(0)})} \bigg )^{1/2}
J_{1/2}(\pi s_1 \rho (x_1^{(0)} - x_2^{(0)})) \Big ) \Big ] \nonumber \\ &&
\quad + \Big ( \pi i \rho (\bar{x}_1^{(0)}+ \bar{x}_2^{(0)}) +
(2 \pi \rho)^2 {\tau \over 2} \Big )
\bigg ( {s_1 \over 2 \pi \rho (x_1^{(0)} - x_2^{(0)})} \bigg )^{1/2}
J_{1/2}(\pi s_1 \rho (x_1^{(0)} - x_2^{(0)})) \bigg \},
\end{eqnarray}
where $\bar{x}_1^{(0)} := x_1^{(0)} - x$, $\bar{x}_2^{(0)} := x_1^{(0)} - x$.
Making use of the formula
\begin{equation}
J_{1/2}(z) = \Big ( {2 \over \pi z} \Big )^{1/2} \sin z,
\end{equation}
summing over $\kappa_1$ and substituting the result in (\ref{fin})
gives that in the thermodynamic limit
\begin{eqnarray}\label{b2}\lefteqn{
\rho_{(2,1)}^T(x_1^{(0)},x_2^{(0)};x;\tau)  = 
2 \rho^3 \int_0^\infty ds \,
\cos \pi s \rho (\bar{x}_1^{(0)}+\bar{x}_2^{(0)})
e^{-(2 \pi \rho)^2 (s^2 + s)\tau / 2}} \nonumber \\
&& + 2 \rho^3 \int_0^\infty ds \,
\Big ( {\sin \pi s \rho (x_1^{(0)} - x_2^{(0)}) \over 
\pi \rho (x_1^{(0)} - x_2^{(0)}) } \Big ) 
\Big ( \pi \rho (\bar{x}_1^{(0)} + \bar{x}_2^{(0)})
\sin \pi s \rho (\bar{x}_1^{(0)}+ \bar{x}_2^{(0)})  \qquad \qquad
\qquad 
\nonumber \\
&& \quad  +
(2 \pi \rho)^2 {\tau \over 2} 
\cos  \pi s \rho (\bar{x}_1^{(0)}+ \bar{x}_2^{(0)}) \Big )
e^{-(2 \pi \rho)^2 (s^2 + s)\tau / 2}.
\end{eqnarray}

As an analytic check on (\ref{b2}), consider the limit
$x_2^{(0)} \to \infty$. Rewriting the trigonometric products
as sums shows that
\begin{equation}\label{lb}
\lim_{x_2^{(0)} \to \infty}
\rho_{(2,1)}^T(x_1^{(0)},x_2^{(0)};x;\tau) = \rho \rho_{(1,1)}^T(x_1^{(0)};
x;\tau)
\end{equation}
as expected.

A  case of special interest  is when $x_1^{(0)} =
x_2^{(0)}$, so that two particles coincide in the initial state.
Making use of the formula
\begin{equation}\label{jo}
P_n^{(\gamma,\gamma)}(1) = \Big ( {n + \gamma \atop n} \Big )
\end{equation}
in (\ref{3.1}) and then proceeding as in the derivation of (\ref{b2})
we have that in the finite system
\begin{eqnarray}\label{b3}
\rho_{(2,1)}^T(x_1^{(0)},x_1^{(0)};x;\tau)&  = &
{2(N-1)(N-2) \over L^3}
{\rm Re}
\sum_{\kappa_1 = 1}^\infty e^{-2 \pi i \kappa_1 \rho (x_1^{(0)} - x)/N}
e^{-(2 \pi \rho / N)^2(\kappa_1^2 + (N - 1) \kappa_1) \tau / 2} \nonumber \\
&& \times \Big (
(\kappa_1 +1) - \kappa_1  e^{-2 \pi i  \rho (x_1^{(0)} - x)/N}
e^{-(2 \pi \rho /N)^2(N - 2) \tau / 2} \Big ).
\end{eqnarray}
It is easy to see that in the thermodynamic limit (\ref{b3})
agrees with (\ref{b2}) with $x_1^{(0)} =
x_2^{(0)}$, thus providing another check on the latter result.
 We can also compare the correlations obtained from the
analytic formula (\ref{b3}) with those obtained empirically from
parameter-dependent Hermitian random matrices with the initial
condition of a doubly degenerate
zero eigenvalue. The parameter-dependent matrices are
constructed in an analogous way to that described at the end of the
previous section. Thus the initial matrix $H^{(0)}$ is
again diagonal and real, with the elements $h_{jj}^{(0)}$, $j > 2$,
chosen with the same Gaussian distribution as before, while the
elements $h_{11}$ and $h_{22}$ are set equal to zero. The matrix
$H$
is chosen so that the diagonal entries
have mean $e^{-t} h_{jj}^{(0)}$ and standard deviation
$(\sqrt{N}/\pi)(1 - e^{-2t})^{1/2}$, $t:= \pi^2 \tau /(2N)$,
while the independent off diagonal elements (real and
imaginary parts of the upper triangular elements)
have mean zero and standard deviation equal to $1/\sqrt{2}$ of that of
the diagonal elements.  Again one feature of this prescription is that
the theoretical density at the origin is unity. 

\begin{figure}
\epsfxsize=10cm
\centerline{\epsfbox{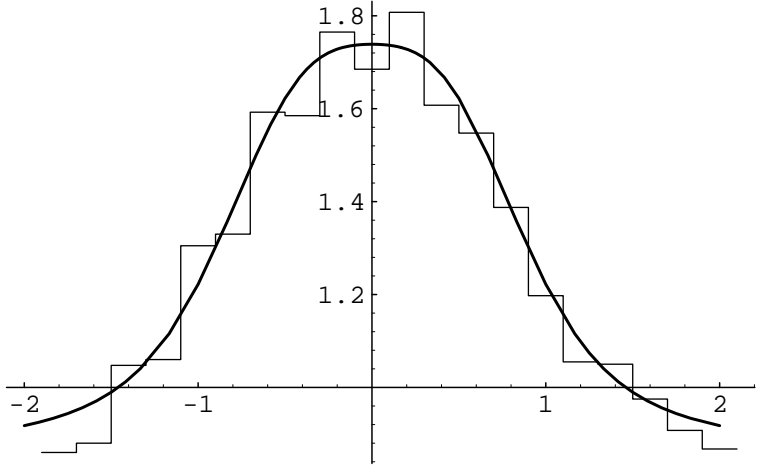}}
\caption{\label{f.2} Comparison between the empirical density for
2,000 Gaussian Hermitian parameter-dependent random matrices 
of dimension $13 \times 13$ with
$t = .05$ and a doubly degenerate zero eigenvalue initially, and
the theoretical value computed from (\ref{b3}). }
\end{figure}

In Figure \ref{f.2}
we give a comparative plot of the empirical density obtained from
generating the eigenvalues of such matrices for a certain choice
of $N$ and $t$ with the corresponding theoretical formula (\ref{b3}).

\vspace{.5cm}

\subsection{$\beta = 4$}
Here the formulas (\ref{u}) and (\ref{kpr}) give
\begin{equation}\label{3u}
u_\kappa \Big |_{\kappa_2=0}= {1 \over L}, \quad
u_\kappa \Big |_{\kappa_2=1}= -{2 \over L}\Big ( 1 - {1 \over \kappa_1 + 2}
\Big ), \quad
u_\kappa \Big |_{\kappa_2=2}= {1 \over L}\Big ( 1 - {2 \over \kappa_1 + 1}
\Big ).
\end{equation}
These values suggest we write the corresponding integrand for the sum over
$\kappa_1$ in the form of a difference approximation to the second
derivative. Doing this, and making use of (\ref{pas}) we find that in the
thermodynamic limit
\begin{eqnarray}\label{bd4}
\rho_{(2,1)}^T(x_1^{(0)},x_2^{(0)};x;\tau) & = &
2 \rho^3 \pi^{1/2} {\rm Re} \int_0^\infty ds \,
e^{-2 \pi i \rho (\bar{x}_1^{(0)} + \bar{x}_2^{(0)})s}
e^{-(2 \pi \rho)^2(s^2 + 2s) \tau / 4} e^{-(2 \pi \rho)^2 s \tau /2} \nonumber
\\ && \times
\Big ( {\partial^2 \over \partial s^2} f(s) -
(\tau / 2) (2 \pi \rho)^2 f(s) + 2 {\partial \over \partial s}
\Big ({f(s) \over s} \Big ) \Big )
\end{eqnarray}
where
\begin{equation}
f(s) := s^{1/2} \bigg ( {1 \over 2 \pi 
\rho (\bar{x}_1^{(0)} - \bar{x}_2^{(0)})} \bigg )^{3/2}
J_{3/2}\Big (\pi \rho  (\bar{x}_1^{(0)} - \bar{x}_2^{(0)}) s \Big )
e^{\pi i \rho (\bar{x}_1^{(0)} + \bar{x}_2^{(0)})s}
e^{(2 \pi \rho)^2 s \tau /2}.
\end{equation}

Regarding checks on this formula,  from the  
large-$x$ expansion
 $J_{3/2}(x) \sim
-(2/(\pi x))^{1/2}\cos x$
it is straightforward to show that the limiting
behaviour (\ref{lb}) is exhibited. Furthermore, use of (\ref{jo})
as well as (\ref{3u}) in (\ref{fin}) gives that in the finite
system with $x_1^{(0)} = x_2^{(0)}$,
\begin{eqnarray}\label{bd5}
\rho_{(2,1)}^T(x_1^{(0)},x_2^{(0)};x;\tau)  = 
{\rho^3 \over 3N} {\rm Re}
\sum_{\kappa_1=1}^\infty e^{-(2 \pi \rho / N)^2(\kappa_1^2 +
2(N - 1) \kappa_1) \tau / 4} e^{-2 \pi i \kappa_1 \rho \bar{x}_1^{(0)}/N}
\Big ( (\kappa_1+3)(\kappa_1+2) \nonumber \\
- 2(\kappa_1+1)^2  e^{-(2 \pi \rho / N)^2(1 + 2(N-3))\tau/4}
e^{2 \pi i \rho x_1^{(0)}/N} + (\kappa_1 - 1)\kappa_1
e^{-(2 \pi \rho / N)^2(4 + 4(N-3))\tau/4}e^{4 \pi i \rho x_1^{(0)}/N}
\Big ). \nonumber \\
\end{eqnarray}
In the thermodynamic limit this agrees with (\ref{bd4}) in the case
$x_1^{(0)} = x_2^{(0)}$ (the formula $J_{\nu}(x) \sim x^\nu/(2^\nu
\Gamma(1 + \nu))$ as $x \to 0$ is required in the checking).

\subsection{General $0 < \beta < 2$}
In the cases $\beta = 2$ and $\beta = 4$ we have seen that cancellation
takes place within the summand of (\ref{fin}) due to the
factor $(-\beta/2)_{\kappa_2}$ having varying sign. This is again true
for general $0 < \beta < 2$. Thus for $\kappa_2=0$, 
$(-\beta/2)_{\kappa_2} = 1$, while for $\kappa_2 \ge 1$,
$(-\beta/2)_{\kappa_2} = -(\beta/2) \Gamma(\kappa_2 - \beta / 2)/
\Gamma(1 - \beta / 2) < 0$. However in the case of general
$0 < \beta < 2$ the factor $(-\beta/2)_{\kappa_2}$ is non-zero for
all $\kappa_2 \ge 0$, so the method of grouping together terms
for cancellation used in the cases $\beta = 2$ and $\beta = 4$ is no longer
applicable.

To gain some insight into the necessary grouping, let us consider the
behaviour of $u_\kappa$, as specified by (\ref{u}) with $\ell(\kappa)=2$,
for large $\kappa_1,\kappa_2$. Use of (\ref{kpr}) and Stirling's
formula shows
\begin{eqnarray}\label{u3}
u_\kappa & = & - {(\beta / 2) \over L}
{(\kappa_1 + \kappa_2) (\kappa_1 - 1)! \over \kappa_2! (\kappa_1 -
\kappa_2)!} {\Gamma(\kappa_2 - \beta/2) \over \Gamma(1 - \beta/2)}
{\Gamma(\beta /2 + \kappa_1 - \kappa_2 + 1) \over 
\Gamma (\beta /2 + \kappa_1+ 1)} \nonumber \\
& \sim & -{(\beta/2) \over L \Gamma(1 - \beta/2)}
{(\kappa_1 - \kappa_2)^{\beta / 2} \over
\kappa_1^{1 + \beta/2} \kappa_2^{1 + \beta / 2}}.
\end{eqnarray}
Writing this in terms of continuous variables $s_1:=\kappa_1/N$,
$s_2 := \kappa_2/N$ we see that (\ref{u3}) is non-integrable at
$s_1 =0$ and $s_2=0$. It is this singularity that we must cancel out
using the $\kappa_2=0$ term.

We proceed as follows. Define
\begin{equation}\label{v}
v_{(\kappa_1,\kappa_2)}
 = {(\kappa_1 + \kappa_2) (\kappa_1 - 1)! \over L (\kappa_1 -
\kappa_2)!} {\Gamma(\beta /2 + \kappa_1 - \kappa_2 + 1) \over 
\Gamma (\beta /2 + \kappa_1+ 1)}
\end{equation}
so that
\begin{equation}\label{defg}
u_\kappa = g_{\kappa_2}v_{(\kappa_1,\kappa_2)}, \qquad g_{\kappa_2} :=
 - {(\beta / 2) \Gamma (\kappa_2 - \beta / 2) \over
\Gamma (1 - \beta / 2) \kappa_2!}
\end{equation}
and
\begin{eqnarray}\label{r21a}\lefteqn{
\rho_{(2,1)}^T(x_1^{(0)},x_2^{(0)};x;\tau) =
{N(N-1) \over L^2}} \nonumber \\&& \times 2 {\rm Re} \Big (
\sum_{\kappa_1 = 1}^\infty v_{(\kappa_1,0)} B_{(\kappa_1,0)}(
z_1^{(0)},z_2^{(0)}) +
\sum_{\kappa_1 \ge \kappa_2 \ge 1}^\infty g_{\kappa_2}
v_{(\kappa_1,\kappa_2)}B_{(\kappa_1,\kappa_2)}(
z_1^{(0)},z_2^{(0)}) \Big ),
\end{eqnarray}
where
\begin{equation}
B_{(\kappa_1,\kappa_2)}(
z_1^{(0)},z_2^{(0)}) := 
\Big ( P_\kappa^{(2/\beta)}(z_1^{(0)},z_2^{(0)}) \Big )^*
e^{2 \pi i (\kappa_1 + \kappa_2) x / L} e^{-\tau (E_\kappa - E_0)/\beta}.
\end{equation}
Now write
\begin{equation}\label{gG}
g_{\kappa_2} = G_{\kappa_2 + 1} - G_{\kappa_2}, \qquad
G_0 := 0, \: G_1 := 1.
\end{equation}
From the general formula
\begin{equation}
\sum_{k=0}^N \Big ( a_k(b_{k+1} - b_k) + b_{k+1} (a_{k+1} - a_k) \Big )=
a_{N+1} b_{N+1} - a_0 b_0,
\end{equation}
we see that (\ref{r21a}) can be rewritten
\begin{eqnarray}\label{ut}\lefteqn{
\rho_{(2,1)}^T(x_1^{(0)},x_2^{(0)};x;\tau)  = 
{N(N-1) \over L^2}} \nonumber \\&&  \times 2 {\rm Re} \Big (
\sum_{\kappa_1 \ge \kappa_2 \ge 1}^\infty 
G_{\kappa_2} \Big ( v_{(\kappa_1,\kappa_2)}B_{(\kappa_1,\kappa_2)}(
z_1^{(0)},z_2^{(0)}) - v_{(\kappa_1,\kappa_2-1)}B_{(\kappa_1,\kappa_2-1)}(
z_1^{(0)},z_2^{(0)}) \Big ) \nonumber \\
&& + \sum_{\kappa_1 = 1}^\infty G_{\kappa_1+1} v_{(\kappa_1,\kappa_1)}
B_{(\kappa_1,\kappa_1)}(z_1^{(0)},z_2^{(0)}) \Big ).
\end{eqnarray}

The utility of (\ref{ut}) is that the summand is well behaved in the
large $\kappa_1$, $\kappa_2$ limit. To see this, note from the definition
(\ref{defg}) that for large $\kappa_2$,
\begin{equation}
g_{\kappa_2} \sim - {(\beta / 2) \over \Gamma (1 - \beta / 2)}
{1 \over \kappa_2^{1 + \beta / 2}},
\end{equation}
and so (\ref{gG}) is asymptotically satisfied with
\begin{equation}\label{G}
G_{\kappa_2} \sim {1 \over \Gamma (1 - \beta / 2)} {1 \over \kappa_2^{\beta /
2}}.
\end{equation}
This has an integrable singularity at the origin, as distinct from the
non-integrable singularity in (\ref{u3}). Making use of the
asymptotic formulas (\ref{G}) and (\ref{pas}) and noting from
(\ref{v}) that
\begin{equation}
v_\kappa \sim {\rho \over N} (\kappa_1 + \kappa_2) |\kappa_1 -
\kappa_2|^{\beta / 2} {1 \over \kappa_1^{1+\beta / 2}},
\end{equation}
we see from (\ref{ut}) that in the thermodynamic limit
\begin{eqnarray}\label{ut1}
\rho_{(2,1)}^T(x_1^{(0)},x_2^{(0)};x;\tau) & = &
{\rho^3 \pi^{1/2} \over (\beta/2)^2\Gamma(-\beta/2)}
\bigg ( {1 \over 2 \pi \rho (\bar{x}_1^{(0)} - \bar{x}_2^{(0)})}
\bigg )^{(\beta - 1)/2} \int_0^\infty ds_1 \int_0^\infty ds_2 \,
\Big ( {1 \over s_1 s_2} \Big )^{\beta / 2} \nonumber \\
&& \times {\partial^2 \over \partial s_1   \partial s_2}
\bigg ( (s_1 + s_2)|s_1 - s_2|^{(\beta + 1)/2}
J_{(\beta - 1)/2}\Big ( |s_1 - s_2| \pi 
\rho (\bar{x}_1^{(0)} - \bar{x}_2^{(0)}) \Big ) \nonumber \\
&& \times \cos \Big (\pi \rho (\bar{x}_1^{(0)} + \bar{x}_2^{(0)})
(s_1 + s_2) \Big ) e^{-(2 \pi \rho)^2(s_1^2 + s_2^2 + (\beta /2)(s_1 +
s_2))\tau / \beta} \bigg ),
\end{eqnarray}
where an integration by parts in $s_1$ has been carried out.

\section{Discussion}
\setcounter{equation}{0}
\subsection{Asymptotics}
In the guise of the Calogero-Sutherland model, the density-density
correlation $\rho_{(1,1)}^T(x_1^{(0)};x;\tau)$ for the Dyson
Brownian motion model in the case that the initial state equals
the equilibrium state (in.$=\,$eq.) is known for all rational
$\beta$ \cite{Ha95} (see \cite{FJ97} or \cite{Fo98} for the explicit
connection between the correlations of the two models). For
$\beta / 2 = p/q$ ($p$ and $q$ relatively prime) its 
value in the thermodynamic limit is given in terms of a
$(p+q)$-dimensional Dotsenko-Fateev-type integral. This contrasts with
the simple one-dimensional integral (\ref{rms1}) for the same
correlation with Poisson initial conditions.

Although the expression for $\rho_{(1,1)}^T(x_1^{(0)};x;\tau)$ 
in the case in.$=\,$eq.~is complicated, its non-oscillatory
leading order
large $|x_1^{(0)} - x|$ and/or $\tau$ expansion is very simple,
being given by \cite{Ha95,FZ96}
\begin{equation}\label{4.1}
\rho_{(1,1)}^T(x_1^{(0)};x;\tau) \Big |_{{\rm in.}={\rm eq.}}
\sim {4 \rho^2 \over \beta}
{\rm Re} \bigg ( {1 \over ({1 \over 2} \tau (2 \pi \rho)^2 + 2 \pi i 
\rho \bar{x}_1^{(0)})^2 } \bigg ).
\end{equation}
Fixing $\tau$, (\ref{4.1}) shows that the leading non-oscillatory
portion of $\rho_{(1,1)}^T(x_1^{(0)};x;\tau)$ falls off as
$-1/(\beta (\pi \bar{x}_1^{(0)})^2)$, while with $ \bar{x}_1^{(0)}$
fixed the decay $1/(\beta (\pi \rho)^2 \tau)^2$ is exhibited. Note
that the decay in $\bar{x}_1^{(0)}$ is independent of the density.

Let us compare (\ref{4.1}) with the corresponding asymptotic
expansion in the case of Poisson initial conditions (in.$=\,$Poi.).
By rewriting the cosine term as a complex exponential, linearizing
the exponent about $s=0$, and changing variables we see 
from (\ref{rms1}) that
\begin{equation}\label{4.2}
\rho_{(1,1)}^T(x_1^{(0)};x;\tau) \Big |_{{\rm in.}={\rm Poi.}}
\sim 2 \rho^2 {\rm Re} \bigg (
{1 \over {\tau \over 2} (2 \pi \rho)^2 - 2 \pi i \rho \bar{x}_1^{(0)}}
\bigg )
\end{equation}
Notice that there is no $\beta$ dependence in this expression.
For $\bar{x}_1^{(0)}$ fixed the decay here is given by
$1/(\pi^2 \tau)$, independent of $\rho$, while for $\tau$ fixed the decay
is $\rho^2 \tau / (\bar{x}_1^{(0)})^2$. These behaviours have distinct
features from those noted above for (\ref{4.1}).

The leading order large $\bar{x}_1^{(0)}$ and/or $\tau$ asymptotic
expansion of (\ref{b2}) ($\beta = 2$ result)
with $\bar{x}_1^{(0)} = \bar{x}_2^{(0)}$
can readily be computed. We find the same behaviour as in
(\ref{4.2}), except that the prefactor $2\rho^2$ is replaced by
$4 \rho^3$. For general $0 < \beta < 2$, setting
$\bar{x}_1^{(0)} = \bar{x}_2^{(0)}$ in (\ref{ut1}), then repeating the
analysis which led to (\ref{4.1}), we find
\begin{equation}\label{4.3}
\rho_{(2,1)}^T(x_1^{(0)};x_1^{(0)};x;\tau)
\sim 4 \rho^3 A(\beta) {\rm Re} \bigg (
{1 \over {\tau \over 2} (2 \pi \rho)^2 - 2 \pi i \rho \bar{x}_1^{(0)}}
\bigg )
\end{equation}
where
\begin{eqnarray}\label{4.4}
A(\beta) &:= &{\pi^{1/2} \over 2^{\beta + 1}(\beta / 2)^2}
{1 \over \Gamma(-\beta/2) \Gamma(\beta/2 + 1/2)} \nonumber \\
&& \times
\int_0^\infty ds_1 \int_0^\infty ds_2
\Big ( {1 \over s_1 s_2} \Big )^{\beta /2}
{\partial^2 \over \partial s_1 \partial s_2} 
\Big ( (s_1 + s_2) |s_1 - s_2|^\beta e^{-(s_1 + s_2)} \Big ).
\end{eqnarray}
The integral in (\ref{4.4}) is evaluated in the Appendix. Subtituting
its value gives $A(\beta) = 1$, so the asymptotic behaviour
found at $\beta = 2$ persists independent of $\beta$
(for $\beta < 2$ at least) analogous to (\ref{4.2}).

\subsection{Fluctuation formulas}
Let us now turn attention to the application of the density-density
correlation $\rho_{(1,1)}^T(x_1^{(0)};\tau)$ in the study of
fluctuation formulas. In general \cite{Sp86} the time
displaced covariance of two linear statistics
\begin{equation}\label{ab}
A_\tau = \sum_{j=1}^N a(x_j(\tau)), \qquad
B_\tau = \sum_{j=1}^N b(x_j(\tau))
\end{equation}
is given in terms of the Fourier transform
\begin{equation}\label{s}
\tilde{S}(k;\tau) := \int_{-\infty}^\infty
\rho_{(1,1)}^T(x_1^{(0)};x;\tau) e^{i \bar{x}_1^{(0)}k} \, d\bar{x}_1^{(0)}
\end{equation}
(assuming a fluid state so $\rho_{(1,1)}^T(x_1^{(0)};x;\tau)$ is a
function of $\bar{x}_1^{(0)} := x_1^{(0)} - x$) according to the
formula
\begin{equation}\label{ab1}
{\rm Cov} (A_0,B_\tau) = \int_{-\infty}^\infty \tilde{a}(k)
\tilde{b}(-k) \tilde{S}(k;\tau)  \, dk.
\end{equation}
Now, from (\ref{rms1}), for the Dyson Brownian motion
model with Poisson initial conditions
\begin{equation}\label{s1}
\tilde{S}(k;\tau) = \rho e^{-k^2 \tau / \beta - \pi \rho |k| \tau},
\end{equation}
so (\ref{ab}) is known explicitly.

Rigorous studies \cite{Sp86} of the infinite Dyson Brownian
motion model for $\beta = 2$ and with in.$=\,$eq., and of the
same model on a circle \cite{Sp98} for general $\beta >0$ have shown that
after 
appropriate scaling the joint distribution of $(A_0,B_\tau)$ is
a Gaussian. In the infinite system the covariance is given by
(\ref{ab1}) with $\tilde{S}(k;\tau)$ replaced by its scaled form,
while for the circle system, a discrete version of (\ref{ab1})
applies. Explicitly, in the infinite system a small parameter
$\epsilon$ is introduced in (\ref{ab}) so that the linear
statistics become
\begin{equation}\label{ab2}
A_\tau = \sum_{j=1}^N a(\epsilon x_j(\tau)), \qquad
B_\tau = \sum_{j=1}^N b(\epsilon  x_j(\tau)).
\end{equation}
Physically this means that the variation of $a$ and $b$ is macroscopic.
Also, $\tau$ is scaled by writing
\begin{equation}\label{ta1}
t = \tau / \epsilon.
\end{equation}
Then it is proved in \cite{Sp86} that in the limit $\epsilon \to
0$ and with $\beta = 2$
the joint distribution of $(A_0,B_t)$ is a Gaussian with
covariance
$$
{1 \over \pi \beta}  \int_{-\infty}^\infty \tilde{a}(k)
\tilde{b}(-k) |k| e^{-|k| \pi \rho t} \, dk.
$$

Also relevant to the present discussion is the corresponding result
for the Brownian dynamics specified by (\ref{fp}) in which the
equilibrium state is compressible (gas with a short range potential).
 Here one introduces the scaled linear statistics by
\begin{equation}\label{ab3}
A_\tau = \epsilon^{1/2} \sum_{j=1}^N a(\epsilon x_j(\tau)), \qquad
B_\tau = \epsilon^{1/2} \sum_{j=1}^N b(\epsilon  x_j(\tau))
\end{equation}
and scales $\tau$ by writing 
\begin{equation}\label{ta2}
t = \tau / \epsilon^2.
\end{equation}
In this setting it is proved in \cite{Sp86a} that the joint distribution
of $(A_0,B_t)$ is a Gaussian with covariance
$$
\chi \int_{-\infty}^\infty \tilde{a}(k)
\tilde{b}(-k)  e^{-k^2   \rho t/ (2 \chi)} \, dk,
$$
where $\chi$ denotes the compressibility. Note that as well as the
different scaling for $\tau$, the linear statistics 
are suppressed by a 
factor $\epsilon^{1/2}$ which is not  required in (\ref{ab2}) for the
Dyson model. This means that in the Dyson model the fluctuations
are naturally suppressed by the long-range nature of the pair
potential.

In the present work we are considering the Dyson model with
Poisson initial conditions. Thus initially the particles are
non-interacting so the system is compressible. A simple calculation
using (\ref{ab1}) and (\ref{rms1}) shows that if we scale
the linear statistics according to (\ref{ab3}) as in a
compressible gas, but scale $\tau$ according to (\ref{ta1}) as
in the Dyson model with in.$=\,$eq., then the covariance in
the limit $\epsilon \to 0$ becomes equal to
\begin{equation}\label{cov2}
\rho \int_{-\infty}^\infty \tilde{a}(k)
\tilde{b}(-k) e^{- \pi \rho |k| t} \, dk.
\end{equation}
We conjecture that in this limit the joint distribution of
$(A_0,B_t)$ is a Gaussian.

Following \cite{Sp98} we can also consider the covariance for the
system on a circle. For this purpose one first scales
$x_1^{(0)}$ and $x$ by multiplying each by $N$, and also scales
$\tau$ by making the replacement (\ref{ta1}). Then, according
to (\ref{rms}), in the thermodynamic limit
\begin{equation}\label{th1}
\rho_{(1,1)}^T(x_1^{(0)},x) \sim {\rho^2 \over N}
\sum_{\kappa_1 = - \infty \atop \kappa \ne 0}^\infty
e^{- 2 (\pi \rho)^2 |\kappa_1| t} e^{2 \pi i \rho (x_1^{(0)}-x) \kappa_1}
\end{equation}
where now $x_1^{(0)},x \in [0,1\rho]$.
In this setting,
analogous to (\ref{ab3}) we choose for the scaled linear
statistics
\begin{equation}\label{ab5}
A_\tau = {1 \over N^{1/2}} \sum_{j=1}^N a(x_j), \qquad
B_\tau = {1 \over N^{1/2}} \sum_{j=1}^N b(x_j).
\end{equation}
The formula for the covariance is then
\begin{equation}\label{cov3}
{\rm Cov}(A_0,B_\tau) = {1 \over N} N^2
\int_0^{1/\rho} dx_1^{(0)} \, a(x_1^{(0)}) \int_0^{1/\rho} dx \,
b(x) \rho_{(1,1)}^T(x_1^{(0)},x),
\end{equation}
where the factor $1/N$ results from the factors of $1/N^{1/2}$ in
(\ref{ab5}), while the factor $N^2$ results from the change of
scale $x_1^{(0)} \to N x_1^{(0)}$, $x \to Nx$. Substituting (\ref{th1})
in (\ref{cov3}) gives that as $N \to \infty$
\begin{equation}\label{cov4}
{\rm Cov}(A_0,B_\tau) = \sum_{\kappa_1 = - \infty \atop \kappa \ne 0}^\infty
a_{\kappa_1} b_{-\kappa_1} e^{-2(\pi \rho)^2 |\kappa_1| t}
\end{equation}
where
$$
a_{\kappa_1} := \rho \int_0^{1/\rho} a(x) e^{2 \pi i \rho x \kappa_1}
\, dx
$$
and similarly the meaning of $b_{-\kappa_1}$. Again we would
expect the joint distribution of $(A_0,B_\tau)$ to be Gaussian.

\subsection*{Acknowledgement}
The financial support of the ARC is acknowledged.

\section*{Appendix}
\renewcommand{\theequation}{A.\arabic{equation}}
\setcounter{equation}{0}
Here we will evaluate the integral
\begin{equation}\label{a.1}
I(\mu,z,\beta) :=
\int_0^\infty ds_1 \int_0^\infty ds_2 \,
(s_1 s_2)^{\mu + 1}
{\partial^2 \over \partial s_1 \partial s_2} 
\Big ( (s_1 + s_2) |s_1 - s_2|^\beta e^{-z(s_1 + s_2)} \Big ),
\end{equation}
which includes the integral in (\ref{4.4}) as a special case.
Assuming temporarily that Re$(\mu) > -1$, integration by parts
gives
\begin{equation}\label{a.2}
I(\mu,z,\beta) = (\mu + 1)^2
\int_0^\infty ds_1 \int_0^\infty ds_2 \,
(s_1 s_2)^{\mu }
(s_1 + s_2) |s_1 - s_2|^\beta e^{-z(s_1 + s_2)}. 
\end{equation}
Furthermore, the term $(s_1 + s_2)$ can be obtained from
$e^{-z(s_1 + s_2)}$ by partial differentiation with respect to $z$.
Doing this, then scaling out the $z$ dependence by changing
variables and computing the derivative gives
\begin{equation}\label{a.3}
I(\mu,z,\beta) = (\mu + 1)^2 (2 + 2 \mu + \beta) z^{-(3 + 2 \mu + \beta)}
L(\mu,\beta)
\end{equation}
where
\begin{eqnarray}\label{a.4}
L(\mu,\beta) & := & \int_0^\infty ds_1 \int_0^\infty ds_2 \,
(s_1 s_2)^{\mu }
(s_1 + s_2) |s_1 - s_2|^\beta e^{-(s_1 + s_2)} \nonumber\\
& = & {\Gamma (1 + \mu) \Gamma (1 + \beta) \Gamma (1 + \mu + \beta / 2)
\over \Gamma (1 + \beta /2)}
\end{eqnarray}
where the second equality in (\ref{a.4})  follows because
$L(\mu,\beta)$ is a two-dimensional example of a well known
limiting case of the $n$-dimensional Selberg integral \cite{Se44},
which in general can be evaluated as a product of gamma functions.

To obtain the integral in (\ref{4.4}) we must put $z=1$ and
$\mu = -1 - \beta /2$. However we see from (\ref{a.4}) that
$L(\mu,\beta)$ diverges with this choice of $\mu$. This is compensated
for by a vanishing factor in (\ref{a.3}), so the correct
procedure is to take the limit $\mu \to -1 - \beta/2$ (an analogous
procedure has been necessary in the evaluation of
similar integrals occuring in random matrix problems
\cite{FZ96}). Doing this, and using the gamma function
identity
$$
2^{2x-1} \Gamma(x) \Gamma (x + 1/2) = \sqrt{\pi} \Gamma (2 x)
$$
shows that
\begin{equation}\label{a.5}
I(-1-\beta/2,1,\beta) = {1 \over \sqrt{\pi}}
2^{\beta + 1} (\beta / 2)^2 \Gamma(-\beta / 2)
\Gamma (\beta /2 + 1/2).
\end{equation}

\end{document}